\documentclass{ws-procs9x6-cpt22}
\begin{document}
\newcommand{\refeq}[1]{(\ref{#1})}
\def\etal {{\it et al.}}
%
\newcommand{\pn}{\ensuremath{\mathrm{p}}}
\newcommand{\pnb}{\ensuremath{\mathrm{\bar{p}}}}
\newcommand{\ksn}{\ensuremath{\mathrm{K_S}}}
\newcommand{\ksln}{\ensuremath{\mathrm{K_{S,L}}}}
\newcommand{\kln}{\ensuremath{\mathrm{K_L}}}
\newcommand{\ksnr}{\ensuremath{\mathrm{K_S^R}}}
\newcommand{\klnr}{\ensuremath{\mathrm{K_L^R}}}
\newcommand{\kkn}{\ensuremath{\mathrm{K}}}
\newcommand{\kknt}{\ensuremath{\mathrm{\widetilde K}}}
\newcommand{\kn}{\ensuremath{\mathrm{K^0}}}
\newcommand{\knb}{\ensuremath{\mathrm{\bar{K}^0}}}
\newcommand{\mn}{\ensuremath{\mathrm{M^0}}}
\newcommand{\mnb}{\ensuremath{\mathrm{\bar{M}^0}}}
\newcommand{\kpp}{\ensuremath{\mathrm{K_+}}}
\newcommand{\kppnn}{\ensuremath{\mathrm{K_{\pm}}}}
\newcommand{\knn}{\ensuremath{\mathrm{K_-}}}
\newcommand{\kpport}{\ensuremath{\mathrm{K_+^{\perp}}}}
\newcommand{\knnpport}{\ensuremath{\mathrm{K_{\mp}^{\perp}}}}
\newcommand{\knnort}{\ensuremath{\mathrm{K_-^{\perp}}}}
\newcommand{\knnmp}{\ensuremath{\mathrm{K_{\mp}}}}
\newcommand{\mpp}{\ensuremath{\mathrm{M_+}}}
\newcommand{\mnn}{\ensuremath{\mathrm{M_-}}}
\newcommand{\kppppm}{\ensuremath{\mathrm{\widetilde K_{\pm}}}}
\newcommand{\kppp}{\ensuremath{\mathrm{\widetilde K_+}}}
\newcommand{\knnp}{\ensuremath{\mathrm{\widetilde K_-}}}
\newcommand{\knt}{\ensuremath{\mathrm{\widetilde K_0}}}
\newcommand{\knbt}{\ensuremath{\mathrm{\widetilde K_{\bar{0}}}}}
\newcommand{\kntt}{\ensuremath{\mathrm{ K_0}}}
\newcommand{\knbtt}{\ensuremath{\mathrm{ K_{\bar{0}}}}}
\newcommand{\bn}{\ensuremath{\mathrm{B^0}}}
\newcommand{\bbn}{\ensuremath{\mathrm{B}}}
\newcommand{\bbnd}{\ensuremath{\mathrm{B_d}}}
\newcommand{\bnb}{\ensuremath{\mathrm{\bar{B}^0}}}
\newcommand{\kknx}{\ensuremath{\mathrm{K_X}}}
\newcommand{\kkny}{\ensuremath{\mathrm{K_Y}}}
\newcommand{\kkone}{\ensuremath{\mathrm{K_1}}}
\newcommand{\kktwo}{\ensuremath{\mathrm{K_2}}}
\newcommand{\kknxb}{\ensuremath{\mathrm{\bar{K}_X}}}
\newcommand{\kknyb}{\ensuremath{\mathrm{\bar{K}_Y}}}
\newcommand\trule{\rule{0pt}{2.6ex}}

\def\CP       {\ensuremath{\mathrm{CP}}}
\def\CPT       {\ensuremath{\mathrm{CPT}}}
\def\C       {\ensuremath{\mathrm{C}}}
\def\P       {\ensuremath{\mathrm{P}}}
\def\T       {\ensuremath{\mathrm{T}}}
\def\SS    {\ensuremath{\mathrm{S}}}
\def\K       {\ensuremath{K}}
\def\B       {\ensuremath{B}}

\title{Entanglement, CPT and neutral kaons}

\author{Antonio Di Domenico}

\address{Dipartimento di Fisica, Sapienza Universit\`a di Roma and
INFN Sezione di Roma \\
P.\ le A.\ Moro 2, I-00185, Rome, Italy}
%

\begin{abstract}
Direct tests of \T, \CP, \CPT\ symmetries in transitions
using entangled neutral kaons produced at a $\phi$-factory
are briefly reviewed.
%
\end{abstract}

\bodymatter

\section{Introduction}
%
Neutral kaons produced at a $\phi$-factory combine their 
peculiar
flavour oscillations, charge-parity (\CP) and time-reversal (\T) violation,
into an Einstein-Podolsky-Rosen (EPR) entangled system, revealing
surprising features.\cite{didohand,ktviol,kcptviol,kloeqm2,future}
Here 
the possibility of exploiting 
their entanglement 
to make direct tests of \T, \CP, \CPT\ symmetries
is briefly reviewed.

\section{Direct test of 
discrete symmetries with neutral kaons}
\par
In order 
to implement direct tests of \T, \CP, \CPT\  symmetries in transitions, the preparation of the initial kaon state
is performed exploiting the 
entanglement of $\kn\knb$ pairs produced at a $\phi$-factory.\cite{didohand,ktviol,kcptviol}
In this way, taken a kaon transition process as a reference,
the exchange of 
{\it in} and {\it out} states 
required for a genuine 
test 
involving an anti-unitary transformation implied by time-reversal,
can be easily implemented.
In fact,
the initial kaon pair produced in $\phi\rightarrow \kn\knb$ decays
can be rewritten in terms of any pair of orthogonal states:
\begin{eqnarray}
  |i \rangle   =  \frac{1}{\sqrt{2}} \{ |\kn \rangle |\knb \rangle - 
 |\knb \rangle |\kn \rangle
\} 
\label{eq:state1}
   =  \frac{1}{\sqrt{2}} \{ |\kpp \rangle |\knn \rangle - 
 |\knn \rangle |\kpp \rangle
\label{eq:state3}
\}~.
\end{eqnarray}  
Here 
the states $|\knn\rangle$, $|\kpp\rangle$ are defined as 
the states which cannot decay into pure $\CP=\pm1$ final states,
$\pi\pi$ 
or $3\pi^0$, respectively.\cite{ktviol,kcptviol}
%
%
%
The condition of orthogonality $\langle\knn|\kpp\rangle=0$, 
corresponds to assume negligible direct \CP\ (or \CPT) violation contributions, while 
the $\Delta S=\Delta Q$ rule is also assumed, so that the two flavor orthogonal eigenstates $|\kn\rangle$ and $|\knb\rangle$ are identified by the charge of the lepton in semileptonic decays.
%
\par
Thus, exploiting the perfect anticorrelation of the states implied 
by Eq.~\refeq{eq:state3},
 it is possible to have a 
\textquotedblleft flavor-tag\textquotedblright or a 
 \textquotedblleft \CP-tag\textquotedblright,
i.e.~to infer the flavor (\kn or \knb) or the \CP\ (\kpp or \knn) state
of the still alive kaon by observing a specific flavor decay ($\pi^+\ell^-\nu$
or  $\pi^-\ell^+\bar{\nu}$) or CP decay ($\pi\pi$ or $3\pi^0$) of the other (and first decaying) kaon in the pair.
Then the decay of the surviving kaon into a semileptonic ($\ell^+$ or $\ell^-$), $\pi\pi$ or $3\pi^0$ final state, filter the kaon final state as a flavor or \CP\ 
state.
%
\par
In this way one can identify 
a reference transition (e.g. $\kn \to \knn$)
and its symmetry conjugate (e.g. the \CPT-conjugated $\knn \to \knb$),
and directly compare them
through
the 
corresponding 
ratios of probabilities.
The observable ratios 
for the various symmetry tests can be defined as follows:\cite{didoactapp}
%
%
\begin{eqnarray}
R_{2,\T}^{\rm{exp}}(\Delta t) &\equiv&
\frac{  I(\ell^-,3\pi^0;\Delta t)}
{ I(\pi\pi,\ell^+;\Delta t)} \cdot \frac{1}{D_{\CPT}}
\nonumber\\
&=&\left(1-4\Re \epsilon +4\Re x_+ + 4 \Re y \right) 
\nonumber\\
& & \times 
\left| 1+ 
\left(
2\epsilon
+\epsilon^{\prime}_{3\pi^0} +\epsilon^{\prime}_{\pi\pi}
\right)
e^{-i(\lambda_S-\lambda_L)\Delta t} \right|^2 
~,
\label{ratio2texp}
\end{eqnarray}
\begin{eqnarray}
R_{4,\T}^{\rm{exp}}(\Delta t) &\equiv&
\frac{  I(\ell^+,3\pi^0;\Delta t)}
{ I(\pi\pi,\ell^-;\Delta t)}   \cdot \frac{1}{D_{\CPT}}
\nonumber\\
&=&\left(1+4\Re \epsilon +4\Re x_+ - 4 \Re y \right) 
\nonumber\\
& & \times \left| 1-
\left(
2\epsilon
+\epsilon^{\prime}_{3\pi^0} +\epsilon^{\prime}_{\pi\pi}
\right)
e^{-i(\lambda_S-\lambda_L)\Delta t} \right|^2
~,
\label{ratio4texp}
\end{eqnarray}
\begin{eqnarray}
R_{2,\CP}^{\rm{exp}}(\Delta t) &\equiv&
\frac{  I(\ell^-,3\pi^0;\Delta t)}
{ I(\ell^+,3\pi^0;\Delta t)}
\nonumber\\
&=&\left(1-4\Re \epsilon_S 
-4\Re x_- + 4 \Re y \right) 
\nonumber\\
& & \times \left| 1+ 
2\left(
\epsilon_S 
+\epsilon^{\prime}_{3\pi^0} 
\right)
e^{-i(\lambda_S-\lambda_L)\Delta t} \right|^2 ~,
\label{ratio2cpexp}
\end{eqnarray}
\begin{eqnarray}
R_{4,\CP}^{\rm{exp}}(\Delta t) &\equiv&
\frac{  I(\pi\pi,\ell^+;\Delta t)}
{ I(\pi\pi,\ell^-;\Delta t)}   
\nonumber\\
&=&\left(1+4\Re \epsilon_L 
-4\Re x_- - 4 \Re y \right) 
\nonumber\\
& & \times \left| 1-
2\left(
\epsilon_L
+\epsilon^{\prime}_{\pi\pi}
\right)
e^{-i(\lambda_S-\lambda_L)\Delta t} \right|^2 ~,
\label{ratio4cpexp}
\end{eqnarray}
\begin{eqnarray}
R_{2,\CPT}^{\rm{exp}}(\Delta t) &\equiv&
\frac{  I(\ell^-,3\pi^0;\Delta t)}
{ I(\pi\pi,\ell^-;\Delta t)} \cdot \frac{1}{D_{\CPT}}
\nonumber\\
&=& \left(1-4\Re \delta+4\Re x_+ - 4 \Re x_-\right) 
\nonumber\\
& & \times 
\left| 1+ 
\left(
2\delta 
+\epsilon^{\prime}_{3\pi^0} -\epsilon^{\prime}_{\pi\pi}
\right)
e^{-i(\lambda_S-\lambda_L)\Delta t} \right|^2
~,
\label{ratio2cptexp}
\end{eqnarray}
\begin{eqnarray}
R_{4,\CPT}^{\rm{exp}}(\Delta t) &\equiv&
\frac{  I(\ell^+,3\pi^0;\Delta t)}
{ I(\pi\pi,\ell^+;\Delta t)} \cdot \frac{1}{D_{\CPT}}
\nonumber\\
&=& \left(1+4\Re \delta+4\Re x_+ + 4 \Re x_-\right) 
\nonumber\\ 
& & \times 
\left| 1-
\left(
2\delta 
+\epsilon^{\prime}_{3\pi^0} -\epsilon^{\prime}_{\pi\pi}
\right)
%
%
%
e^{-i(\lambda_S-\lambda_L)\Delta t} \right|^2
~,
\label{ratio4cptexp}
\end{eqnarray}
where
$I(f_1,f_2;\Delta t)$
are the double decay rates
into 
decay products
$f_1$ 
and $f_2$ 
as a function of the difference of kaon decay times $\Delta t$,\cite{didohand,ktviol,kcptviol} with
$f_1$ occurring before $f_2$ decay for $\Delta t >0$, and viceversa for $\Delta t<0$.
$D_{\CPT}$ is a constant factor that can be determined from measurable branching fractions and lifetimes of \ksln\ states~\cite{kcptviol}.
For $\Delta t=0$ one has by construction, within our assumptions, $R_{i,\SS}^{\rm{exp}}(0)=1$ (with $\SS=\T,\CP$, or $\CPT$, and $i=2,4$), and
the measurement of any deviation from the prediction $R_{i,\SS}^{\rm{exp}}(\Delta t)=1$
imposed by the symmetry invariance 
is a 
direct signal of the symmetry violation built in the time 
evolution of the system.
The following double ratios independent of the factor $D_{\CPT}$ can also be defined:
\begin{eqnarray}
\label{eq:dratiot}
DR_{\T,\CP} (\Delta t) &\equiv & \frac{R_{2,\T}^{\rm{exp}}(\Delta t)}{ R_{4,\T}^{\rm{exp}}(\Delta t)} \equiv \frac{R_{2,\CP}^{\rm{exp}}(\Delta t)}{R_{4,\CP}^{\rm{exp}}(\Delta t)} 
\nonumber\\
&=& \left(1-8\Re \epsilon+8\Re y \right) 
\nonumber\\ 
& & \times 
\left| 1+2
\left(
2\epsilon
+\epsilon^{\prime}_{3\pi^0} +\epsilon^{\prime}_{\pi\pi}
\right)
%
%
%
e^{-i(\lambda_S-\lambda_L)\Delta t} \right|^2
~,
\\
\label{eq:dratiocpt}
DR_{\CPT} (\Delta t) & \equiv & \frac{R_{2,\CPT}^{\rm{exp}}(\Delta t)}{R_{4,\CPT}^{\rm{exp}}(\Delta t)} 
\nonumber\\
&=& \left(1-8\Re \delta-8\Re x_-\right) 
\nonumber\\ 
& & \times 
\left| 1+2
\left(
2\delta 
+\epsilon^{\prime}_{3\pi^0} -\epsilon^{\prime}_{\pi\pi}
\right)
%
%
%
e^{-i(\lambda_S-\lambda_L)\Delta t} \right|^2
~.
\end{eqnarray}  
The r.h.s. of Eqs.~\refeq{ratio2texp}-\refeq{eq:dratiocpt} 
is evaluated to first order in small parameters
and for not too large negative $\Delta t$;
$\epsilon$ and $\delta$ are the usual \T\ and \CPT\ violation parameters in the neutral
kaon mixing, respectively, and $\epsilon_{S,L}=\epsilon\pm\delta$ the \CP\ impurities in the 
physical states \ksn\ and \kln;
the small parameter $y$ describes a possible \CPT\ violation in the $\Delta S = \Delta Q$ semileptonic decay amplitudes, while
$x_+$ and $x_-$ describe $\Delta S \neq \Delta Q$ semileptonic decay amplitudes
with \CPT\ invariance and \CPT\ violation, respectively.
Therefore the r.h.s. of Eqs.~\refeq{ratio2texp}-\refeq{eq:dratiocpt} 
shows
the effect of
symmetry 
violations only in the
the effective Hamiltonian description of the neutral kaon system according to the
Weisskopf-Wigner approximation,
without the presence of other possible sources of symmetry violations.
The small spurious effects due to the release of our assumptions are also shown, including possible $\Delta S=\Delta Q$ rule violations ($x_+,x_-\neq0$)
and/or direct \CP\ and/or \CPT\ violation effects ($\epsilon^{\prime}_{3\pi^0},\epsilon^{\prime}_{\pi\pi},y\neq0$). In particular
the $\epsilon^{\prime}$ effects are fully negligible in the asymptotic region $\Delta t \gg \tau_S$.
\par
The KLOE-2 collaboration has recently analysed a data sample corresponding to an integrated luminosity $L=1.7~\hbox{fb}^{-1}$ collected 
at the DA$\Phi$NE $\phi$-factory, and measured all eight observables defined in Eqs.\ \refeq{ratio2texp}-\refeq{eq:dratiocpt} in the asymptotic region $\Delta t \gg \tau_S$ (with positive $\Delta t$).\cite{kloetcpcpt} These results constitute the first direct tests of \T\ and \CPT\ symmetries in kaon transitions.
The \T\ and \CPT\ observables have been measured with a precision of few percent, showing no evidence of symmetry violations, 
while \CP\ violation has been observed with more than $5\sigma$ significance with the observable ratio \refeq{ratio4cpexp}.
 \par
The double ratio $DR_{\CPT}$
in  Eq.\ \refeq{eq:dratiocpt} 
in the asymptotic regime 
appears one of 
the best observable for testing \CPT, free from approximations and model independent~\cite{kcptviol}.
Thus it seems best suited to extend the \CPT\ tests to
the general framework of the Standard Model Extension (SME) for \CPT\ and Lorentz symmetry breaking,\cite{datatables,kost1,kost2,kost3,agnes,kost4,agnesralf}
e.g. by studying the \CPT\ observables 
as a function of 4-momenta of the kaons.\cite{kloecpt2013}
\par Finally it is worth noting that the full exploitation of time correlations of the two entangled kaons, not only {\it from past to future},
as in the above mentioned direct \T, \CP, and \CPT\ tests, but also {\it from future to past}, as described in Ref.\ \refcite{future}
to post-tag special kaon states,
might lead to the identification of new observables suitable to further extend these symmetry tests.

\section*{Acknowledgments}
The author 
would like to warmly thank R.~Lehnert for the invitation to this very interesting edition of the meeting.

\end{document}